\documentclass[12pt,nofootinbib]{article}

\usepackage{graphicx}
\usepackage{amssymb}
\usepackage{amsmath}
\usepackage{color}

\IfFileExists{srcltx.sty}{\usepackage[active]{srcltx}}

\begin{document}
\renewcommand{\thefootnote}{\#\arabic{footnote}}
\newcommand{\rem}[1]{{\bf [#1]}} \newcommand{\gsim}{ \mathop{}_
{\textstyle \sim}^{\textstyle >} } \newcommand{\lsim}{ \mathop{}_
{\textstyle \sim}^{\textstyle <} } \newcommand{\vev}[1]{ \left\langle
{#1} \right\rangle } \newcommand{\bear}{\begin{array}} \newcommand
{\eear}{\end{array}} \newcommand{\bea}{\begin{eqnarray}}
\newcommand{\eea}{\end{eqnarray}} \newcommand{\beq}{\begin{equation}}
\newcommand{\eeq}{\end{equation}} \newcommand{\bef}{\begin{figure}}
\newcommand {\eef}{\end{figure}} \newcommand{\bec}{\begin{center}}
\newcommand {\eec}{\end{center}} \newcommand{\non}{\nonumber}
\newcommand {\eqn}[1]{\beq {#1}\eeq} \newcommand{\la}{\left\langle}
\newcommand{\ra}{\right\rangle} \newcommand{\ds}{\displaystyle}
\newcommand{\red}{\textcolor{red}} 
\def\SEC#1{Sec.~\ref{#1}} \def\FIG#1{Fig.~\ref{#1}}
\def\EQ#1{Eq.~(\ref{#1})} \def\EQS#1{Eqs.~(\ref{#1})} \def\lrf#1#2{
\left(\frac{#1}{#2}\right)} \def\lrfp#1#2#3{ \left(\frac{#1}{#2}
\right)^{#3}} \def\GEV#1{10^{#1}{\rm\,GeV}}
\def\MEV#1{10^{#1}{\rm\,MeV}} \def\KEV#1{10^{#1}{\rm\,keV}}
\def\REF#1{(\ref{#1})} \def\lrf#1#2{ \left(\frac{#1}{#2}\right)}
\def\lrfp#1#2#3{ \left(\frac{#1}{#2} \right)^{#3}} \def\OG#1{ {\cal
O}(#1){\rm\,GeV}}

\begin{titlepage}

\begin{flushright}
ICRR-Report 635-2012-24 \\
IPMU12-0198
\end{flushright}

\begin{center}

\vskip 1.2cm

{\Large\sf
High Scale SUSY Breaking From Topological Inflation 
}

\vskip 1.2cm

Keisuke Harigaya$^a$, Masahiro Kawasaki$^{b,a}$
and
Tsutomu T. Yanagida$^a$

\vskip 0.4cm

{\it$^a$Kavli Institute for the Physics and Mathematics of the Universe,
The University of Tokyo, 5-1-5 Kashiwanoha, Kashiwa, Chiba 277-8583, Japan}\\
{\it$^b$Institute for Cosmic Ray Research, The University of Tokyo,
5-1-5 Kashiwanoha, Kashiwa, Chiba 277-8582, Japan}

\date{\today}

\vskip 1.2cm

\begin{abstract}
The recently observed mass $\sim$ 125 GeV for the Higgs boson suggests a
 high-energy scale SUSY breaking, above $O(10)$ TeV. It is, however,
 very puzzling why nature chooses such a high energy scale for the SUSY
 breaking, if the SUSY is a solution to the hierarchy problem. We show
 that the pure gravity mediation provides us with a possible solution to
 this puzzle if the topological inflation is the last inflation in the
 early universe.
We briefly discuss a chaotic inflation model in which a similar solution
 can be obtained.

\end{abstract}


\end{center}
\end{titlepage}

\baselineskip 6mm

\section{Introduction}

The ATLAS and CMS collaborations recently discovered a standard-model like Higgs boson of mass about 125 GeV \cite{Higgs}. This observed Higgs mass, together with non-discovery of superpartners at LHC, suggests that the supersymmetry (SUSY) breaking scale is much higher 
than we expected, say above $O(10)$ TeV \cite{Okada:1990vk}. However, if
the SUSY is a solution to the hierarchy problem and hence its breaking
is biased toward low energy scales, a crucial question naturally arises
; why does nature choose such a high energy scale for the SUSY breaking
\cite{Yanagida:2010zz}? We show, in this letter, that the pure gravity mediation
model recently proposed to explain the 125 GeV Higgs mass
\cite{Ibe:2011aa} (for a similar model, see also \cite{Hall:2012zp}) provides us with a possible explanation for the high
scale SUSY breaking if the topological inflation is the last inflation in the early universe.

It is believed that our universe experienced the quasi-exponential expansion 
called inflation \cite{Guth:1980zm} at its very early stage. 
Inflation makes our universe homogeneous and flat, which solves conceptual 
problems of big-bang cosmology, and also dilutes harmful relics like monopoles.  
Furthermore, quantum fluctuations of the inflaton ($=$ a scalar field that drives 
inflation) become classical by the cosmic expansion during inflation and result in 
density perturbations of the universe \cite{Mukhanov:1981xt}. 
Inflation predicts  nearly scale-invariant, adiabatic and gaussian density 
perturbations, which are consistent with the recent observations of the cosmic 
microwave radiation (CMB)~\cite{Komatsu:2010fb}. 
Thus, the inflationary universe successfully describes our 
universe.

However, most inflation models have so called initial value 
problem \cite{Linde:2005ht}, that is, they  require tuning for the initial conditions of 
the inflaton and other relevant fields. 
Among many models, chaotic inflation~\cite{Linde:1983gd} and 
topological inflation~\cite{Linde:1994hy} are free from 
the initial value problem. 
It is well known that chaotic inflation occurs naturally from large field 
fluctuations at the Planck time. 
In topological inflation models, some discrete symmetry is spontaneously broken 
and topological defects (domain walls) are formed in the early universe. 
If the scalar field forming the defects has the vacuum expectation 
value larger than the Planck scale ($M_{\rm pl} \simeq 2.4\times 10^{18}$~GeV), 
the region inside the domain wall undergoes inflation. 
Since the defect formation is inevitable, inflation takes place naturally 
as long as the universe lives until the start of inflation. 
The longevity of the universe is not a problem in the open universe. 
Furthermore, open universes are likely created through tunneling in
quantum cosmology \cite{Coule:1999wg}. 

We show, in this letter, that there is an upper bound of the reheating temperature,
$T_R \lsim 10^{10}$ GeV, if the topological inflation takes place in the early universe. 
In the pure gravity mediation model, the wino is the LSP and it is the
unique candidate of dark matter (DM) in the universe. The number density
of the wino is almost proportional to the reheating temperature $T_R$ as
long as its mass is lower than $1$ TeV \cite{Ibe:2011aa}. We  thus
obtain  a  lower bound of the wino mass to explain the observed DM
density as $m_{{\rm wino}} \gsim 200$ GeV. The lower bound of the wino
mass is translated to the lower bound on the gravitino mass, $m_{3/2}
\gsim {\cal O}(10)$ TeV in the pure gravity mediation model, implying
scalar masses $\gsim {\cal O}(10)$ TeV. Thus, there is a cosmological reason why the SUSY breaking scale is higher than  $O(10)$ TeV in the pure gravity mediation model if the topological inflation is the last inflation in our universe. 
We briefly note, in the last section, that a similar conclusion can be
obtained in a chaotic inflation model.


\section{Topological Inflation Model}

%

It was pointed out long time ago~\cite{Izawa:1998rh,Kawasaki:2000tv}
that the topological inflation takes place for the following simple
super potential and K\"{a}hler potential with $U(1)_R\times Z_2$ symmetry in the supergravity:
\begin{eqnarray}
   W & = & v^2 X (1-\sum_{n}\frac{g_{2n}}{(2n)!}\phi^{2n}) 
   \label{eq:super}\\
   K & = & |X|^2 + |\phi |^2 + k_1 |X |^2|\phi |^2 + \frac{k_2}{4}|X |^4,
   \label{eq:kahler}
\end{eqnarray}
where $g_{2n}~(n=1,2,..), ~k_1$ and $k_2$ are coupling constant.
We take a unit of the Planck scale $M_{\rm pl}=1$, here and hereafter.
We have assumed that the expansions of super and K\"{a}hler potentials
in the fields, $X$ and $\phi$, are well defined as long as the absolute
values of the fields are within the Planck scale.

Thus, it is reasonable to consider all constants $g_{2n},~ k_1$ and $k_2$ are at most $O(1)$. For our analysis we neglect the higher order terms with $n\geq 2$ in the super potential, for simplicity. This may be valid as long as the
expansion in $\phi$ converges sufficiently fast. In the followings, we
assume this property about the expansion in the inflaton $\phi$. 
We have neglected possible higher order terms already in 
the K\"{a}hler potential in Eq. (\ref{eq:kahler}). We assume this model
throughout this letter and consider that the main conclusion of this
letter does not change even if we adopt another model for the
topological inflation.
$g\equiv g_{2}$ is chosen to be real and positive by a phase
rotation of $\phi$.

$X$ and $\phi$ have $U(1)_R$ charges 2 and 0, respectively.
We also assume that $X$ is even and $\phi$ is odd under the $Z_2$, which
is essential for the topological inflation to take place.
The potential has a vacuum
(see Eq.~(\ref{eq:super}) with $n=1$),
\begin{equation}
   \vev{X} = 0, ~~~~\vev{\phi} = \sqrt{\frac{2}{g}}.
\end{equation}
As shown in Ref.~\cite{Kawasaki:2000tv}, the topological inflation takes 
place if $\vev{\phi} \gsim 1/\sqrt{2}$ ($g \lesssim 4$).
Since $\langle \phi \rangle \lsim \cal{O}$$(1)$ for $g={\cal O}(1)$,
it is consistent with our assumption of neglecting higher order terms in
the super and the K\"{a}hler potential. 

The scalar potential derived from Eqs.~(\ref{eq:super}) with $(n=1)$ and (\ref{eq:kahler}) 
is, for $|X|$ and $|\phi|\ll1$,
\begin{equation}
   V = v^4 |1-\frac{g}{2}\phi^2|^2 \left[1+ (1-k_1)|\phi |^2 -k_2|X |^2)\right].
\end{equation}
The $X$ field quickly settles down to the origin if $k_2 \lesssim -1$, 
so hereafter we take $X=0$.
We can identify the inflaton field as the real part of $\phi$.
Using $\varphi = \sqrt{2}{\rm Re}(\phi)$, the potential is rewritten 
for $\varphi \ll 1$ as
\begin{equation}
   V \simeq v^4 -\frac{1}{2}(g +k_1-1)v^4 \varphi^2
   \equiv v^4 -\frac{1}{2}\kappa v^4\varphi^2
\end{equation}
From this potential we obtain the $e$-folds $N$ as
\begin{equation}
   N = \int_{\varphi_N}^{\varphi_f} d\varphi \frac{V}{V'} 
   \simeq \frac{1}{\kappa}\ln\left(\frac{\varphi_f}{\varphi_N}\right),
\end{equation}
where $\varphi_f$ is the field value of $\varphi$ at the end of inflation. 
The slow roll parameters are given by
\begin{eqnarray}
   \epsilon & = & \frac{1}{2}\frac{V'^2}{V^2} 
   = \frac{1}{2} \kappa^2 \varphi^2 \ll \eta\\[0.5em]
   \eta  & = & \frac{V''}{V} = -\kappa 
\end{eqnarray}
For the inflation to produce the observed curvature perturbation, 
the inflaton potential satisfies
$V^{3/2}(\varphi_N)/V'(\varphi_N) \simeq 5\times 10^{-4}$ 
for $N=50-60$, which leads to
\begin{equation}
  v \simeq 0.023 \sqrt{\kappa} e^{-\kappa N/2} \simeq 10^{-3}
  \label{eq:normalization}
\end{equation}
for $\kappa=0.01$, where we have used $\varphi_f \simeq 1$.
The inflaton mass is 
$ m_{\phi}\simeq v^2\sqrt{2g}\simeq 10^{13}~{\rm GeV}$
for $g\simeq 1$.

\begin{figure}[t]
\centering
\includegraphics[width = 12cm]{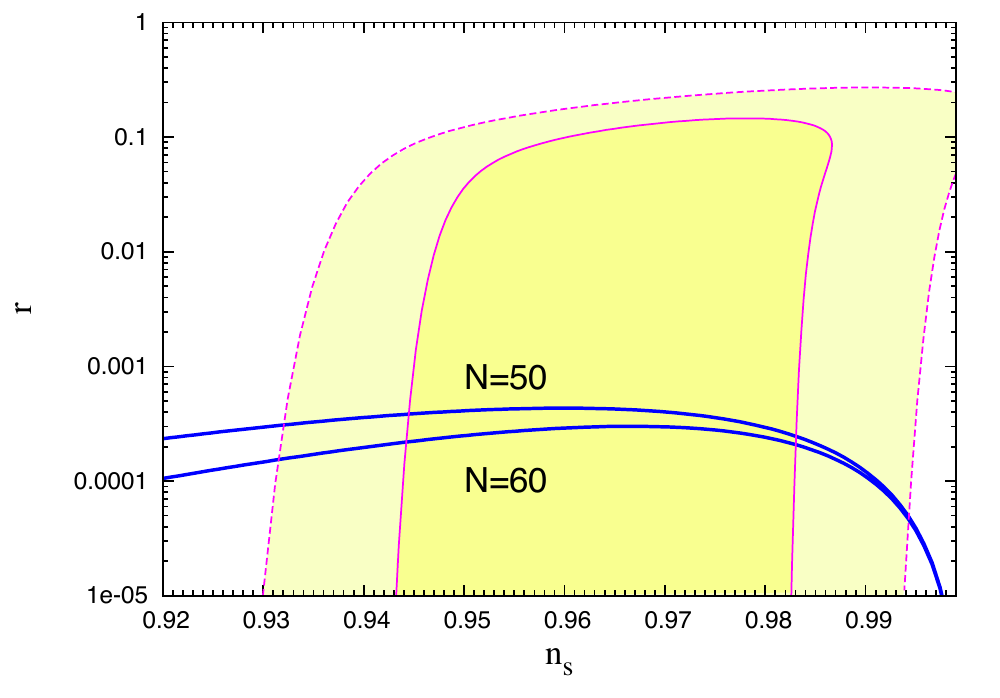}
\caption{
	Prediction by the topological inflation in the $r$-$n_s$ plane.
	We also show the constraint from WMAP~\cite{Komatsu:2010fb}.}
\label{fig:ns-r}
\end{figure}

The spectral index $n_s$ and tensor to scalar ratio $r$ are given by
\begin{eqnarray}
   n_s & = & 1 -6\epsilon + 2\eta = 1-2\kappa , 
   \label{eq:index}\\[0.5em]
   r & = & 16\epsilon = 8\kappa^2 e^{-2\kappa N}.
\end{eqnarray}
Thus, $r$ is written as a function of $N_s$,
\begin{equation}
   r = 2(1-n_s)^2 \exp[-(1-n_s)N]
\end{equation}
The prediction of the topological inflation is shown in Fig.~\ref{fig:ns-r} 
together with WMAP 7year constraint~\cite{Komatsu:2010fb}.
For the spectral index $n_s \simeq 0.94-0.98$ which is consistent with 
WMAP 7 year data, $r \simeq (3-4)\times 10^{-4}$.
Therefore, the tensor mode that the topological inflation produces will be 
not detected even in the future satellite experiments such as
CMBPol~\cite{Baumann:2008aq} and LiteBIRD~\cite{LiteBIRD}.

We find that the value of the coupling constant $g$ is further
restricted. If $g$ is too small, the vacuum $\vev{\phi}=\sqrt{2/g}$
is far above one. On the other hands, a factor of exp$(|\phi|^2)$ in
the inflaton potential lifts up the potential above
$\phi\simeq 1$. Therefore, the potential becomes a old-inflation type and
the topological inflation fails. Now let us estimate the lower bound for
$g$. The potential of the inflaton field is, for $X=0$,
\begin{eqnarray}
 V&=&
v^4{\rm exp}(\frac{\varphi^2}{2})(1+\frac{k_1}{2}\varphi^2)^{-1}(1-\frac{g}{4}\varphi^2)^2\nonumber\\
&\simeq& v^4{\rm
 exp}(\frac{\varphi^2}{2})(1+\frac{1-g}{2}\varphi^2)^{-1}(1-\frac{g}{4}\varphi^2)^2
\label{eq:potential accurate}
\end{eqnarray}
In the second line we have imposed $k_1+g\simeq 1$ which is required by the
observation of the spectral index.
The potential becomes a old-inflation type if there exists a point such
that $\partial V/\partial \varphi=0$ for $0<\varphi<2/\sqrt{g}$. By a
simple algebraic calculation, we find that the condition
\footnote{Higher order terms in K\"{a}hler and super potentials may
relax the constraints (\ref{eq:g lower bound}) and (\ref{eq:g upper bound}).}  
\begin{eqnarray}
g>2-\sqrt{2}\simeq 0.59
\label{eq:g lower bound}
\end{eqnarray}
is required.
On the other hand, if g is too large, $k_1\simeq 1-g$ is negatively
large. From Eq.~(\ref{eq:potential accurate}), one can see that the
inflaton potential becomes singular before $\varphi$ reaches the vacuum. In
order to avoid such behavior, the condition
\begin{eqnarray}
g<2
\label{eq:g upper bound}
\end{eqnarray}
is required.
Eqs.~(\ref{eq:g lower bound}) and (\ref{eq:g upper bound})
are consistent with our assumption that $g$ is of order one.

Now, let us estimate the reheating temperature after the inflation. 
There are four types of $R$ and $Z_2$ invariant interactions which
contribute dominantly to the inflaton decay. One originates from the
K\"{a}hler potential of the form
\begin{equation}
   K= \sum_{n+m={\rm 2}} \frac{c'''_{nm}}{n!m!}\phi^n\phi^{*m}|\Psi|^2,
\end{equation}
where $\Psi$ is any field which is lighter than the half of the inflaton mass and
$c'''_{nm}$ is a coupling constant of order one.
However, with this type of interactions, the matrix element of the inflaton
decay is proportional to square of the mass of $\Psi$ for any $n$
and $m$. Since the
mass of $\Psi$ should be smaller than the half of the inflaton mass,
decay width is suppressed in comparison with the mode described
below at least by the factor of 16. Therefore, we ignore the contribution from this type of
interactions.

The other dominant interaction originates from the K\"{a}hler potential of the form
\begin{equation}
K= c''\phi\phi^{*}H_uH_d
\end{equation}
where $H_u$ and $H_d$ are the up and down type Higgs field and $c''$ is
the coupling constant of order one.
This term is allowed since $H_uH_d$ has a vanishing $U(1)_R$ charge in
the pure gravity mediation model in order to achieve an appropriate value for the $B\mu$
and $\mu$ term \cite{Ibe:2011aa}.
This K\"{a}hler potential leads to the inflaton interaction with the Higgs
scalar as
\begin{eqnarray}
 {\cal L}_{\rm int}=c'' \phi \partial_{\mu}(H_u H_d)\partial^{\mu}\phi^* + h.c.
\end{eqnarray}
The decay rate of the
inflaton to Higgs bosons is given by
\begin{equation}
  \Gamma_{\phi\rightarrow H_u H_d} = \frac{|c''|^2 }{8\pi}\langle
   \phi\rangle^2 m_\phi^3,
\label{eq: decay rate Higgs}
\end{equation}
where we have used the inflaton mass $m_\phi = v^2\sqrt{2g}$. The decay
rate into Higgsinos is the same as this value.

There also exist interactions originate from the super potential.
The small value of the $v^2$ required for the topological inflation in
Eq.~(\ref{eq:normalization}) is considered as a result of some new symmetry　
breaking, otherwise it should be ${\cal O}(1)$.
For instance, consider a parity under which both of $X$ and $v^2$
transform as odd, then $v^2X$ is a completely neutral under all
symmetries except for $U(1)_R$ (the $R$ charge of $X$ is two).
The small value of $v$ is regarded as a small breaking of the parity.
Thus, a
term
\begin{eqnarray}
 W = c' v^2XH_uH_d
\end{eqnarray}
is allowed in the super potential.
This super potential leads
to the inflaton interaction with the Higgs bosons as
\begin{eqnarray}
 {\cal L}_{\rm int}=\frac{g}{2}v^4 \phi^{*2} c'H_uH_d + h.c.
\end{eqnarray}
The decay rate due
to this operator is given by
\begin{eqnarray}
 \Gamma_{\phi\rightarrow H_uH_d}&=&\frac{|c'|^2}{8\pi} g^2
  \frac{v^8}{m_\phi}\vev{\phi}^2\nonumber\\
&=&\frac{|c'|^2}{32\pi}m_\phi^3\vev{\phi}^2.
\end{eqnarray}

The last dominant interactions originate from the gauge kinetic functions:
\begin{equation}
   f =  \frac{\delta_{ab}}{4}(1 + \frac{c}{2}\phi^2)W^a_\alpha W^b_\alpha, 
\end{equation}
where $a, b, \cdots$ are the indices for gauge group and $c$ is a coupling
constant of order one.
This kinetic function leads to inflaton interactions with the gauge fields 
in the standard model as
\begin{equation}
{\cal L}_{\rm int} = -\frac{1}{4}{\rm Re}(\frac{c}{2}\phi^2)
 F_{\mu\nu}^a F^{a\mu\nu}+\frac{1}{8}{\rm Im}(\frac{c}{2}\phi^2) \epsilon^{\mu\nu\rho\sigma}F_{\mu\nu}^a F^{a}_{\rho\sigma},
\end{equation}
where $F_{\mu\nu}^a$ is the field strength of the gauge boson 
$A_{\mu}^a$, and Re and Im denote the real part and the imaginary part, respectively.
Then the decay rate of the inflaton to the gauge fields is given by 
\begin{equation}
  \Gamma_{\phi\rightarrow A^a A^a} = \frac{N_A|c|^2 }{128\pi}\langle
   \phi\rangle^2 m_\phi^3,
\label{eq: decay rate gauge}
\end{equation}
$N_A=1+3+8=12$ is
the number of the gauge fields.
The contribution from the decay into gauginos is exactly the same one.

\begin{figure}[t]
\centering
\includegraphics [width = 11cm]{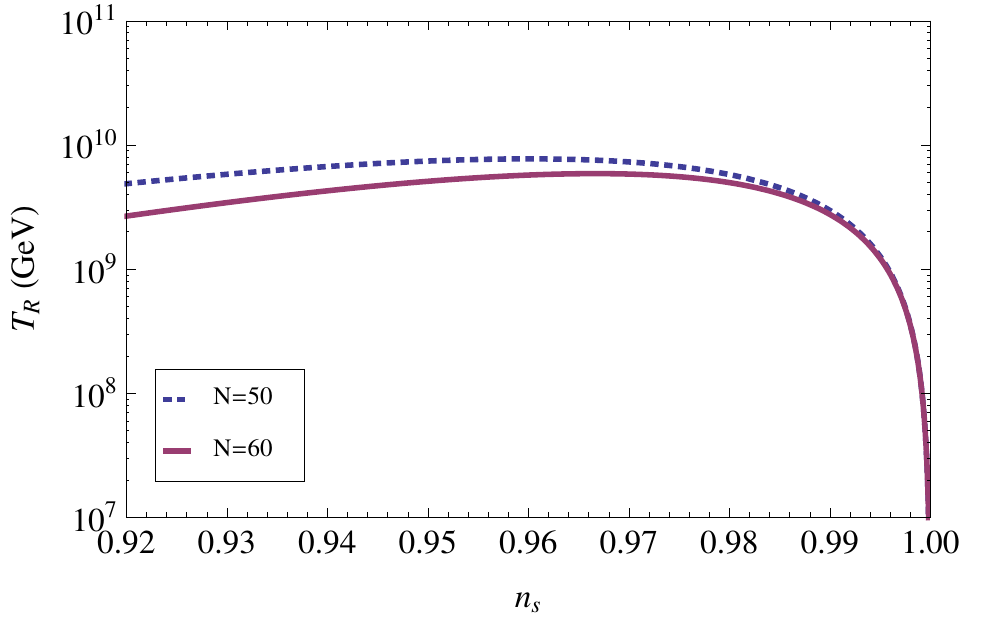}
\caption{
	Reheating temperature for $N=50$(upper curve) and $60$(lower curve). 
	We take $((8|c''|^2+|c'|^2+6|c|^2)/15)^{\frac{1}{2}}g^{1/4}=1$. }
\label{fig:reheating}
\end{figure}

Adding above contributions together, the decay rate of the inflaton
is given by
\begin{eqnarray}
   \Gamma_{\phi} = \frac{8|c''|^2+|c'|^2+6|c|^2 }{32\pi}\langle
   \phi\rangle^2 m_\phi^3,
\label{eq: decay rate}
\end{eqnarray}
From this decay rate we obtain the reheating temperature $T_R$ as 
\begin{equation}
 T_R \simeq 0.25 (\frac{8|c''|^2+|c'|^2+6|c|^2}{15})^{\frac{1}{2}}g^{-\frac{1}{2}} m_{\phi}^{3/2}
   = 0.41 (\frac{8|c''|^2+|c'|^2+6|c|^2}{15})^{\frac{1}{2}}  g^{1/4}v^3 .
\end{equation}
Using Eqs.~(\ref{eq:normalization}) and (\ref{eq:index}), the reheating 
temperature is written as a function of the spectral index $n_s$,
\begin{equation}
   T_R = 4.3\times 10^{12}{\rm GeV} (\frac{8|c''|^2+|c'|^2+6|c|^2}{15})^{\frac{1}{2}} g^{1/4} (1-n_s)^{3/2}
   \exp\left[ -\frac{3}{4}N (1-n_s)\right],
\end{equation}
which is shown in Fig.~\ref{fig:reheating}.
We can see that $T_R \simeq (2-7)\times 10^{9}$~GeV in the present model. 
This relatively low reheating temperature is a consequence of the
$Z_2$ and the $U(1)_R$ symmetry, which is essential for the topological
inflation.
Taking into account ${\cal O}(1)$ ambiguity in the coupling constants
$c$'s, we safely conclude $T_R\lsim10^{10}$ GeV.

\section{Lower Bound on the Gravitino Mass in the Pure Gravity Mediation Model}

In the previous section, we have shown that the maximal reheating temperature
is about $T_R\lsim 10^{10}$ GeV in a reasonable parameter range if the
topological inflation is the last inflation in our universe. Now we are
at the point to show that there is indeed a lower bound on the SUSY
breaking energy scale, $m_{3/2}\gsim {\cal O}(10)$ TeV. 

First of all, the wino is the LSP and the unique candidate for the DM in the pure gravity mediation model \cite{Ibe:2011aa}. The wino has a large annihilation cross section and hence the thermal wino can not provide a sufficient density for the observed DM  as long as the mass of the wino $m_{{\rm wino}} \lsim 1$ TeV. Thus, we must invoke a non thermal wino production in the early universe. The most promising process is the gravitino decay, since the gravitino is much heavier than the wino in the pure gravity mediation model.

Let us discuss the gravitino production in the high-energy thermal
particle scattering in the early universe. Since the production of the gravitino occurs by
Planck suppressed interactions, the production is more effective for
higher temperatures. Therefore, the abundance of
the gravitino is determined by the reheating
temperature and given by \cite{Kawasaki:1994af}
\begin{eqnarray}
 Y_{3/2}\equiv \frac{n_{3/2}}{s}\simeq 2.3 \times
  10^{-12} \times\frac{T_{\rm R}}{10^{10}~{\rm GeV}}.
\label{eq: gravitino abundance}
\end{eqnarray}
Here, $n_{3/2}$ is the number density of the gravitino and $s$
is the entropy density of the universe.

The abundance of the wino from the decay of the gravitino is
the same as Eq.~(\ref{eq: gravitino abundance}), $n_{\rm wino}=n_{3/2}$.
After all, the energy fraction of
the wino dark matter in the present universe $\Omega_{\rm wino,0}$ is
given by
\footnote{For $m_{\rm wino}\gsim 1$ TeV, the
contribution from the thermally produced wino is not negligible.
However, we can safely neglect this contribution in order to discuss the lower bound on the wino mass.}
\begin{eqnarray}
 \Omega_{\rm wino,0} \simeq \frac{m_{\rm wino} Y_{3/2}s_0}{\rho_{\rm cr,
  0}}\simeq 0.12 h^{-2}\frac{m_{\rm wino}}{200~{\rm
  GeV}}\frac{T_{\rm R}}{10^{10}~{\rm GeV}},
\end{eqnarray}
where $s_0\simeq 2.2 \times 10^{-11}~{\rm eV}^3$ is the entropy density
of the present universe, $\rho_{\rm cr,0}\simeq 8.1 h^{-2}\times
10^{-11}~{\rm eV}^4$ is the critical density of the present universe and
$h$ is the scale factor for Hubble constant defined by $H_0= 100 h~{\rm
km~sec^{-1}~Mpc^{-1}}$.

The anthropic bound of the dark matter density is known as
$\Omega_{\rm DM}h^2\gsim 0.1$ \cite{Hellerman:2005yi}.
And as we have seen in the previous section, the reheating temperature
$T_{\rm R}$ is bounded from above as $T_{\rm R} \lsim 10^{10}$ GeV if the
topological inflation is the last inflation in the early universe. Therefore, we
obtain the lower bound on the wino mass as
\begin{eqnarray}
 m_{\rm wino}\gsim 200~{\rm GeV}.
\end{eqnarray}
In the pure gravity mediation model, the wino mass is determined by the
contribution from the anomaly mediation and the threshold correction
by the Higgsino loop. Both contributions are of the same order and
the wino mass has the upper bound as \cite{Ibe:2011aa}
\begin{eqnarray}
 m_{\rm wino}\lsim 10^{-2} m_{3/2}.
\end{eqnarray}
Therefore, we obtain the lower bound for the SUSY breaking scale $m_{\rm
SUSY}$; 
\begin{eqnarray}
 m_{\rm SUSY}~\simeq m_{3/2}> {\cal O}(10)~{\rm TeV}.
\end{eqnarray}

\section{Conclusions and Discussion}

We have discussed the topological inflation in the previous section. However, as pointed out in the introduction, the chaotic inflation \cite{Linde:1983gd} is also interesting since there is no initial value problem.
The chaotic inflation is easily constructed by using a shift symmetry in
supergravity \cite{Kawasaki:2000yn}. The K\"{a}hler potential is a
function of $\phi+\phi^\dagger$  and the super potential is given by
\begin{eqnarray}
 W= mX\phi,
\end{eqnarray}
where $\phi$ is the inflaton. The inflaton $\phi$ can have an interaction in the K\"{a}hler potential,
\begin{eqnarray}
 K= d(\phi +\phi^\dagger)H_uH_d,
\end{eqnarray}
where $d$ is a coupling constant of $O(1)$. It is easy to see that the upper bound of the  reheating temperature $T_R \lsim10^{10}$ GeV can be obtained for $d\lsim O(1)$.

In this letter, we consider relatively high energy scales for the SUSY
breaking. However, there is a very natural parameter region where the
observed DM density is explained by a mixed wino-bino thermal relic DM
of mass ${\cal O}(1)$ GeV and sfermion masses are ${\cal O}(100)$ GeV,
even in the pure gravity mediation model
\cite{Feldstein-Yanagida}. However, this parameter region is excluded by
too much non thermal DM production due to the overproduction of the
gravitino in the inflaton decay \cite{Endo:2007sz}.

We have assumed the $R$ parity conservation throughout this
letter. However, if the $R$ parity is broken, the present argument is
not applicable. The model which connects the SUSY breaking and the
Peccei-Quinn symmetry breaking dynamics \cite{Feldstein-Yanagida}
may be interesting to understand the high scale SUSY breaking, if it is
the case.

\section*{Acknowledgments}
This work is supported by Grant-in-Aid for Scientific research from the
Ministry of Education, Science, Sports, and Culture (MEXT), Japan, No. 14102004 (M.K.), No. 21111006 (M.K.), No.\ 22244021 (T.T.Y.), and also by World Premier International Research Center Initiative (WPI Initiative), MEXT, Japan.
 The work of K.H. is supported in part by a JSPS Research Fellowships for Young Scientists.


\begin{thebibliography}{99}

\bibitem{Higgs}
G. Aad et al. [ATLAS Collaboration], Phys. Lett. B 716, 1 (2012);
S. Chatrchyan et al. [CMS Collaboration], Phys. Lett. B716, 30 (2012).

\bibitem{Okada:1990vk} 
  Y.~Okada, M.~Yamaguchi and T.~Yanagida,
  Prog.\ Theor.\ Phys.\  {\bf 85}, 1 (1991);
  Y.~Okada, M.~Yamaguchi and T.~Yanagida,
  Phys.\ Lett.\ B {\bf 262}, 54 (1991);
  J.~R.~Ellis, G.~Ridolfi and F.~Zwirner,
  Phys.\ Lett.\ B {\bf 257}, 83 (1991);
  H.~E.~Haber and R.~Hempfling,
  Phys.\ Rev.\ Lett.\  {\bf 66}, 1815 (1991).

\bibitem{Yanagida:2010zz} 
  T.~T.~Yanagida and K.~Yonekura,
  Phys.\ Lett.\ B {\bf 693}, 281 (2010)
  [arXiv:1006.2271 [hep-ph]];
  F.~Takahashi and T.~T.~Yanagida,
  Phys.\ Lett.\ B {\bf 698}, 408 (2011)
  [arXiv:1101.0867 [hep-ph]];
  M.~Bose and M.~Dine,
  arXiv:1209.2488 [hep-ph];
  B.~Feldstein and T.~T.~Yanagida,
  arXiv:1210.7578 [hep-ph].

\bibitem{Ibe:2011aa} 
  M.~Ibe and T.~T.~Yanagida,
  Phys.\ Lett.\ B {\bf 709}, 374 (2012)
  [arXiv:1112.2462 [hep-ph]];
  M.~Ibe, S.~Matsumoto and T.~T.~Yanagida,
  Phys.\ Rev.\ D {\bf 85}, 095011 (2012)
  [arXiv:1202.2253 [hep-ph]];
  B.~Bhattacherjee, B.~Feldstein, M.~Ibe, S.~Matsumoto and T.~T.~Yanagida,
  arXiv:1207.5453 [hep-ph].

\bibitem{Hall:2012zp}
  L.~J.~Hall, Y.~Nomura and S.~Shirai,
  arXiv:1210.2395 [hep-ph].

\bibitem{Guth:1980zm} 
  A.~H.~Guth,
  Phys.\ Rev.\ D {\bf 23}, 347 (1981);
see also
  A.~A.~Starobinsky,
  Phys.\ Lett.\ B {\bf 91}, 99 (1980);
  K.~Sato,
  Mon.\ Not.\ Roy.\ Astron.\ Soc.\  {\bf 195}, 467 (1981).

\bibitem{Mukhanov:1981xt} 
  V.~F.~Mukhanov and G.~V.~Chibisov,
  JETP Lett.\  {\bf 33}, 532 (1981)
  [Pisma Zh.\ Eksp.\ Teor.\ Fiz.\  {\bf 33}, 549 (1981)].

\bibitem{Komatsu:2010fb} 
  E.~Komatsu {\it et al.}  [WMAP Collaboration],
  Astrophys.\ J.\ Suppl.\  {\bf 192}, 18 (2011)
  [arXiv:1001.4538 [astro-ph.CO]].

\bibitem{Linde:2005ht} 
  A.~D.~Linde,
  Contemp.\ Concepts Phys.\  {\bf 5}, 1 (1990)
  [hep-th/0503203].

\bibitem{Linde:1983gd} 
  A.~D.~Linde,
  Phys.\ Lett.\ B {\bf 129}, 177 (1983).

\bibitem{Linde:1994hy} 
  A.~D.~Linde,
  Phys.\ Lett.\ B {\bf 327}, 208 (1994)
  [astro-ph/9402031];
  A.~Vilenkin,
  Phys.\ Rev.\ Lett.\  {\bf 72}, 3137 (1994)
  [hep-th/9402085].

\bibitem{Coule:1999wg} 
  D.~H.~Coule and J.~Martin,
  Phys.\ Rev.\ D {\bf 61}, 063501 (2000)
  [gr-qc/9905056];
 L.~Susskind,
 In *Carr, Bernard (ed.): Universe or multiverse?* 247-266
 [hep-th/0302219].


\bibitem{Izawa:1998rh} 
  K.~I.~Izawa, M.~Kawasaki and T.~Yanagida,
  Prog.\ Theor.\ Phys.\  {\bf 101}, 1129 (1999)
  [hep-ph/9810537].

\bibitem{Kawasaki:2000tv} 
  M.~Kawasaki, N.~Sakai, M.~Yamaguchi and T.~Yanagida,
  Phys.\ Rev.\ D {\bf 62}, 123507 (2000)
  [hep-ph/0005073].

  
  
    
   
 
\bibitem{Baumann:2008aq}
  D.~Baumann {\it et al.}  [CMBPol Study Team Collaboration],
  AIP Conf.\ Proc.\  {\bf 1141}, 10 (2009)
  [arXiv:0811.3919 [astro-ph]].

\bibitem{LiteBIRD}
LiteBIRD project, \verb$http://cmb.kek.jp/litebird/index.html$.

\bibitem{Kawasaki:1994af} 
  M.~Kawasaki and T.~Moroi,
  Prog.\ Theor.\ Phys.\  {\bf 93}, 879 (1995)
  [hep-ph/9403364, hep-ph/9403061];
 M.~Kawasaki, K.~Kohri, T.~Moroi and A.~Yotsuyanagi,
 Phys.\ Rev.\ D {\bf 78}, 065011 (2008)
 [arXiv:0804.3745 [hep-ph]].


\bibitem{Hellerman:2005yi} 
  S.~Hellerman and J.~Walcher,
  Phys.\ Rev.\ D {\bf 72}, 123520 (2005)
  [hep-th/0508161];
  M.~Tegmark, A.~Aguirre, M.~Rees and F.~Wilczek,
  Phys.\ Rev.\ D {\bf 73}, 023505 (2006)
  [astro-ph/0511774].


  
\bibitem{Kawasaki:2000yn} 
  M.~Kawasaki, M.~Yamaguchi and T.~Yanagida,
  Phys.\ Rev.\ Lett.\  {\bf 85}, 3572 (2000)
  [hep-ph/0004243].

\bibitem{Feldstein-Yanagida} B. Feldstein and T. T. Yanagida in \cite{Yanagida:2010zz}.

\bibitem{Endo:2007sz} 
  M.~Endo, F.~Takahashi and T.~T.~Yanagida,
  Phys.\ Rev.\ D {\bf 76}, 083509 (2007)
  [arXiv:0706.0986 [hep-ph]].

\end{thebibliography}
\end{document}